\documentclass[11pt]{article}
\usepackage[final]{acl}
\usepackage{microtype}
\usepackage{amsmath}
\usepackage{amssymb}
\usepackage{booktabs}
\usepackage{graphicx}
\usepackage{arydshln}
\usepackage{multirow}
\usepackage{float}
\usepackage{subcaption}
\usepackage{comment}
\usepackage{flushend}

\title{Exploring Multimodal Turn-Taking Cues in Face-to-Face Conversation using Voice Activity Projection}

\begin{document}

\author{
  Willem Berner \\
  Lund University \\
  \texttt{willem.berner@gmail.com}
  \And
  Julio Cesar Cavalcanti \\
  KTH Royal Institute of Technology \\
  \texttt{jcca@kth.se}
  \AND
  Kalle {\AA}str{\"o}m \\
  Lund University \\
  \texttt{karl.astrom@math.lth.se}
  \And
  Gabriel Skantze \\
  KTH Royal Institute of Technology \\
  \texttt{skantze@kth.se}
}
\maketitle

\begin{abstract}
Turn-taking is a fundamental component of spoken interaction, and while humans naturally rely on both verbal and non-verbal signals, dialogue systems usually depend on audio cues alone. This paper investigates whether visual features from face-to-face conversations can enhance turn-taking prediction beyond what is achievable from audio-only. We extend the Voice Activity Projection (VAP) model, a self-supervised transformer-based model for predicting future voice activity, by incorporating visual features extracted from the large-scale Meta Seamless Interaction dataset of dyadic face-to-face conversations. The visual features include gaze direction, head movement, body and hand pose, and facial action units (FAU). For incorporating the visual features, we explore concatenation, cross-attention fusion, delta features, and trainable gating mechanisms. Results show that visual information improves performance over the audio-only baseline, with FAU being significantly more informative than other feature groups. Body and gaze features nevertheless contribute complementary information, as the model combining all features performs best. Furthermore, results indicate that performance on specific tasks varies depending on whether training and test data come from improvised (acted) or naturalistic (non-acted) conversations. 

\end{abstract}

\section{Introduction}

Turn-taking in conversation appears effortless between humans, yet relies on a rich interplay of verbal and non-verbal cues across both visual and auditory modalities \cite{cowley1998,skantze2021,duncan1972signals}. Turn-taking behaviour also varies substantially across individuals and dyads, suggesting that conversational timing is also affected by interaction-level dynamics  \cite{cavalcanti2025}. While telephone conversations rely solely on auditory cues, such as lexical content and prosody, face-to-face interactions additionally incorporate visual cues, including gaze direction and gestures, to coordinate turn transitions \cite{duncan1972signals}. 

\begin{figure}[t]
\centering
\includegraphics[width=0.48\columnwidth]{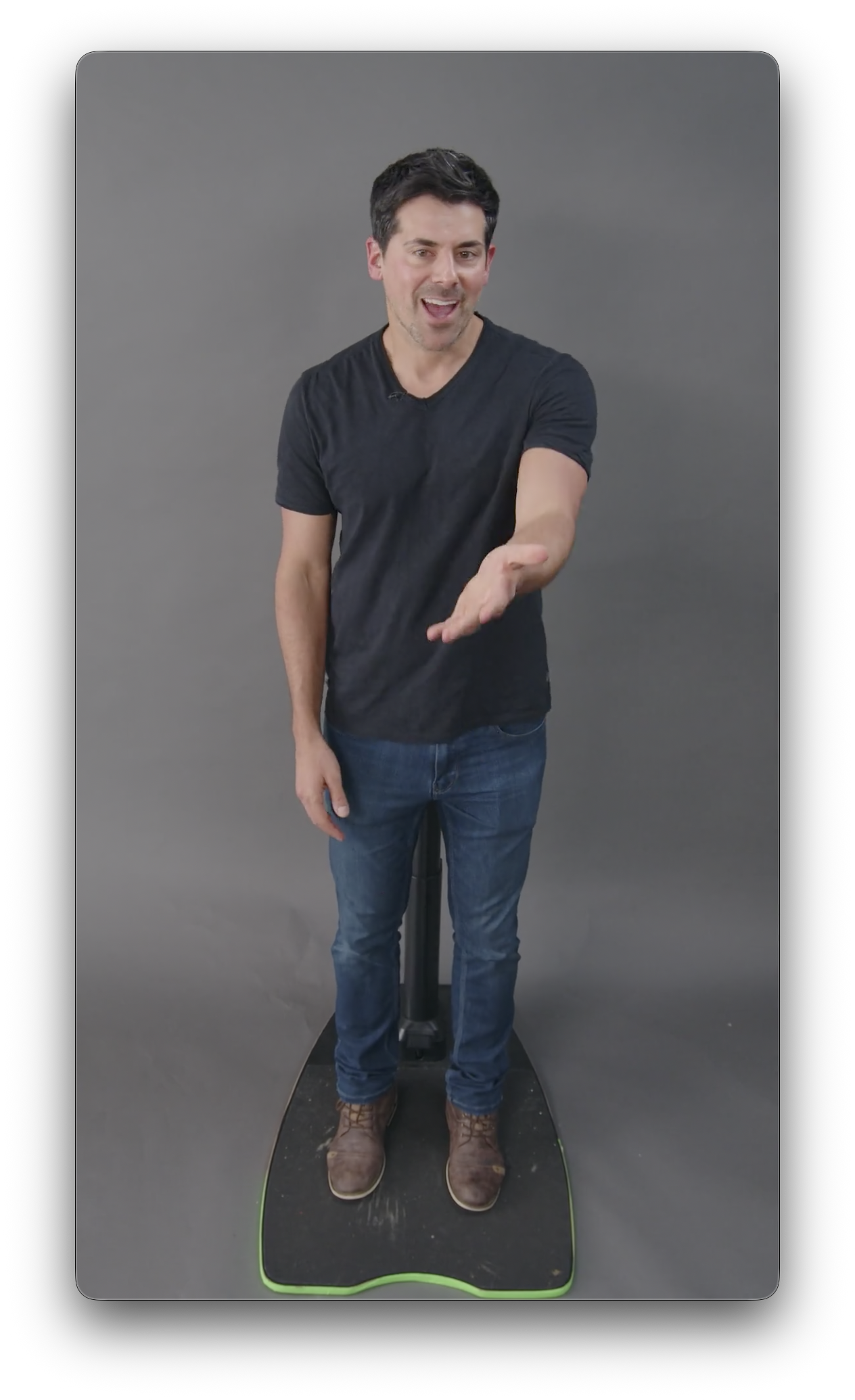}%
\hspace{0.2cm}%
\includegraphics[width=0.48\columnwidth]{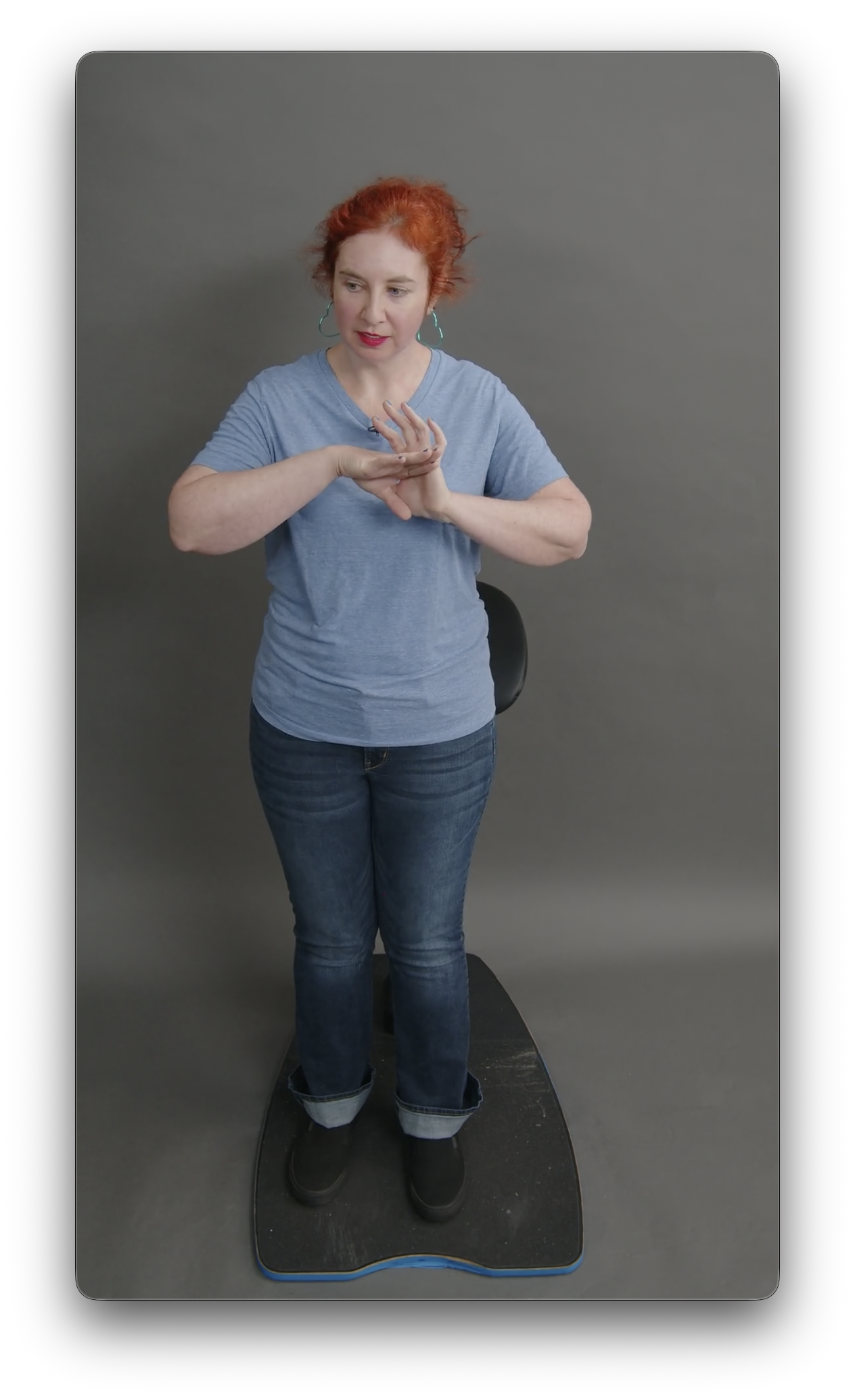}
\caption{Two participants from the Seamless Interaction dataset.}
\label{fig:samples}
\end{figure}

In spoken dialogue systems, turn-taking has traditionally been managed with a silence threshold. While simple, this approach often leads to interruptions or sluggish responses. Transformer-based models have significantly improved prediction of turn-taking, in particular the Voice Activity Projection (VAP) model \cite{ekstedt2022}, which uses self-supervised learning on raw audio to predict which speaker(s) will be active over the next two seconds. VAP has been shown to outperform silence-threshold-based models for turn-taking \cite{skantze2025applying}.  

A natural question is whether visual information can further improve such an audio-only model. Prior work has explored this, but on small or video-conferencing datasets \cite{saga2025,oconnor2025}. Video-conferencing settings differ substantially from face-to-face interaction: turn transitions are shorter and more frequent in person, and gaze estimation is inherently ambiguous on-screen \cite{tian2023, horstmann2022}. Findings from such conditions may therefore not generalise to face-to-face dialogue.

The release of the Meta Seamless Interaction dataset \cite{agrawal2025}, shown in Figure \ref{fig:samples}, provides an opportunity to address this gap. It contains 4000 hours of recorded face-to-face conversations with paired audio, video, and pre-extracted visual features. In this work, we propose an architecture of VAP enhanced with visual features, trained on approximately 400 hours of data. 

Concretely, our main question is: \textit{Can visual cues from face-to-face conversations improve turn-taking prediction beyond audio alone, and, if so, which feature groups or individual features contribute most?}
As a secondary exploratory analysis, we examine predictive performance within and across the corpus’s \textit{improvised} and \textit{naturalistic} subsets. These comprise prompted role-play involving actors and prompted, unscripted interactions between non-actors, respectively. We therefore ask: \textit{How does turn-taking prediction performance vary when models are trained and evaluated within and across the improvised and naturalistic subsets?}

\section{Related Work}

Turn-taking is governed by a systematic organisation of cues that signal when a speaker intends to yield, hold, or take a turn \cite{sacks1974,duncan1972signals}. While prosody and timing are strong predictors \cite{cowley1998,ekstedt_how_2022}, face-to-face interlocutors also rely heavily on visual signals such as gaze and gesture \cite{duncan1972signals,kendon100127,ter_bekke_hand_2024}. Turn-taking dynamics further differ across communication settings: transitions are shorter and overlaps more frequent in person than in mediated, on-screen conversation \cite{tian2023}, motivating the study of visual cues specifically in large-scale face-to-face data.

\citet{ekstedt2022} introduced VAP, a self-supervised framework in which a transformer is trained to predict future binary voice-activity bins for two speakers, operating on raw audio via a frozen CPC encoder \cite{oord2018}. VAP has since been widely adopted for turn-taking prediction, extended to real-time and continuous operation \cite{inoue2024realtime}, as well as multilingual settings \cite{inoue_multilingual_2024}, and forms the basis of our work.

Several studies have extended VAP with visual information. \citet{onishi2023} were the first to do so, augmenting VAP with non-verbal features -- gaze, facial action units (FAU), head pose, and upper-body joints -- on the NoXi corpus of mediated expert-novice interactions. They reported that FAU contributed most to turn-taking prediction. \citet{saga2025} created a similar model that was trained on the French subset of NoXi (7 hours). They replaced hand-crafted action-unit inputs with a pre-trained facial-expression encoder; their models were competitive with, and on shift- and backchannel-prediction better than, prior multimodal baselines, though the small dataset limits generalisation. \citet{oconnor2025} introduced MM-VAP, combining speech with facial expression, head pose, and gaze, and evaluated it on a 710-hour subset of the CANDOR video-conferencing corpus, again finding facial-expression features the most informative and outperforming an audio-only baseline on hold/shift prediction.  

Across these studies, the FAU are consistently the most predictive, but all prior multimodal VAP work is confined to either small corpora \cite{onishi2023,saga2025} or video-conferencing recordings \cite{oconnor2025}. Whether these findings hold for large-scale, face-to-face in-person (co-situated) conversation remains open. We address this using the Meta Seamless Interaction dataset \cite{agrawal2025}, evaluating which visual feature groups improve turn-taking prediction at scale.

Beyond the VAP framework, other work has pursued multimodal turn-taking prediction with independently designed architectures. \citet{wang2024turntaking} combined a neural acoustic model with a large language model, fusing lexical and acoustic representations to predict turn-taking. The experiments demonstrated that this multi-modal fusion of lexical and acoustic information consistently outperforms baseline models relying on a single modality. Their approach, however, is limited to audio and text and does not incorporate visual signals. \citet{lin2025predicting} extended this line of work to a third modality, proposing a tri-modal architecture with text, audio and video, fusing separate pretrained backbones (GPT-2, HuBERT, VideoMAE) through a flexible fusion module. Their model frames the task as discrete classification of keep, turn-taking, and backchannel actions. To support this work, the authors also introduced MM-F2F, a tri-modal (text, audio, video) dataset with both turn-taking and backchannel annotations, comprising over 210 hours of face-to-face conversation.

\section{Data}

\subsection{Seamless Interaction Dataset}

We use the open-source Meta Seamless Interaction dataset \cite{agrawal2025}, which comprises over 4000 hours of face-to-face conversations between two participants. Recordings include two separated mono audio channels and corresponding video files, with pre-extracted visual features provided per frame. The corpus distinguishes improvised role-play involving actors from naturalistic prompted interactions involving non-actors. The two variants are labelled \textit{improvised} and \textit{naturalistic} in the dataset. We acknowledge that naturalistic interaction is not to be treated as equivalent to unelicited everyday conversation. We use a subset of approximately 400 hours total, split into training (328h), validation (41h), and test (41h) sets with a balanced improvised/naturalistic distribution in each split (Table~\ref{tab:splits}). The three sets are conversation-disjoint but not participant- nor dyad-disjoint: dyads in the test set also occur in training and validation.

\begin{table}[htbp]
\centering
\small
\begin{tabular}{lccc}
\toprule
\textbf{Split} & \textbf{Total} & \textbf{Improvised} & \textbf{Naturalistic} \\
\midrule
Train & 328h & 47.3\% & 52.7\% \\
Val   & 41h  & 46.3\% & 53.7\% \\
Test  & 41h  & 46.3\% & 53.7\% \\
\bottomrule
\end{tabular}
\caption{Dataset partitions of our dataset.}
\label{tab:splits}
\end{table}

\subsection{Visual Features}

We use four groups of visual features:

\textbf{Gaze features} ($\mathbb{R}^2$ per frame): pitch and yaw of gaze direction per speaker, extracted by a Meta-internal tool. Values are noisy when eyes are partially occluded.

\textbf{Head features} ($\mathbb{R}^3$ per frame): pitch, yaw, and roll of head orientation.

\textbf{Body and hand features}: Body pose is represented as 21 joints ($b_i \in \mathbb{R}^3$) from the SMPL-H model \cite{romero2017}, estimated via HMR2.0 \cite{goel2023}. Hand pose is represented as 15 joints ($h_i \in \mathbb{R}^3$) per hand, estimated with ViTPose \cite{xu2022}.

\textbf{Facial Action Units (FAU)}: 24 action units from the Facial Action Coding System (FACS) \cite{ekman1978}, each with intensity values $ \in [0,1]$. Gaze-related units are excluded as gaze is already represented separately.

We define the following feature subsets for ablation: \textit{All features} (body, hands, gaze, FAU); \textit{Body features} (body, hands); \textit{Gaze features} (gaze, head); \textit{FAU features} (FAU only). The full visual feature vector has dimension 182.

The visual features are supplied at 30~Hz; we linearly interpolate them to match the baseline's 50~Hz rate.

\section{Methodology}

\begin{figure}[t]
  \centering
  \includegraphics[width=0.48\columnwidth]{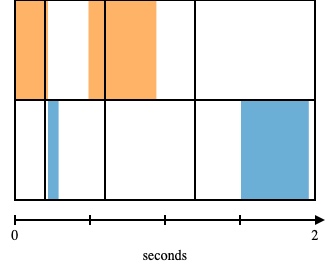}%
  \includegraphics[width=0.48\columnwidth]{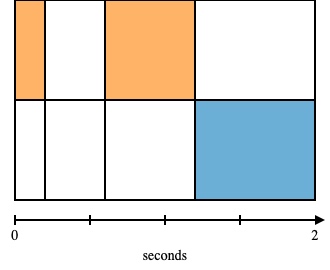}
  \caption{Voice activity before (left) and after binning (right). Each bin is active if the majority of its duration contains speech. Upper row: speaker A; lower row: speaker B.}
  \label{fig:binning}
\end{figure}

\subsection{Baseline: Voice Activity Projection}

The VAP model \cite{ekstedt2022} encodes raw audio via a frozen CPC encoder \cite{oord2018}, producing 256-dimensional latent representations at 50~Hz. For each of the two speakers, a self-attention transformer models within-speaker context, producing representations $X_a$
and $X_b$ for speakers A and B, respectively. This is followed by a bidirectional cross-attention between speakers 

\[
\begin{aligned}
\hat{X}_a &= \text{CrossAttn}(X_a, X_b), \\
\hat{X}_b &= \text{CrossAttn}(X_b, X_a). 
\end{aligned}
\]

All attention operations use four heads. A combinator merges the two enriched representations:
\[
y = \text{GELU}(\text{LayerNorm}(W_a\hat{X}_a + W_b\hat{X}_b)).
\]

The VAP head predicts the probability distribution over all 256 combinations of 8 binary future voice activity bins (4 per speaker, spanning 2 seconds). Bins are spaced as $[(0.0, 0.2); (0.2, 0.6); (0.6, 1.2); (1.2, 2.0)]$ seconds, giving finer resolution for the near future. For a bin to be considered as voice active, the majority of it must be covered (Figure \ref{fig:binning}). The voice activity (VA) head predicts current VA via a sigmoid. The combined training loss is $\mathcal{L} = \mathcal{L}_\text{VAP} + \mathcal{L}_\text{VA}$, with model selection based on $\mathcal{L}_\text{VAP}$ alone. The full architecture of the VAP model is shown in Figure \ref{fig:M0-M5}.

\begin{figure*}[t]
  \centering
  \captionsetup[subfigure]{font=footnotesize}

  \begin{subfigure}{0.3\textwidth}
    \centering
    \setlength{\tabcolsep}{2pt}
    \begin{tabular}{@{}cc@{}}
      \includegraphics[width=0.48\linewidth]{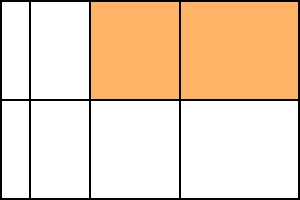} &
      \includegraphics[width=0.48\linewidth]{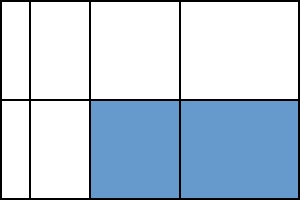} \\
    \end{tabular}
    \caption{Hold vs.\ Shift}
  \end{subfigure}
  \hspace{8pt}
  \begin{subfigure}{0.3\textwidth}
    \centering
    \setlength{\tabcolsep}{2pt}
    \begin{tabular}{@{}cc@{}}
      \includegraphics[width=0.48\linewidth]{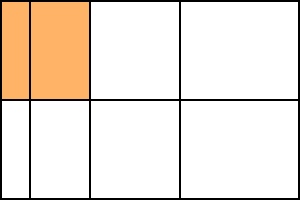} &
      \includegraphics[width=0.48\linewidth]{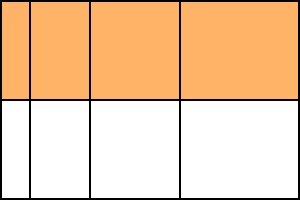} \\
    \end{tabular}
    \caption{Short vs.\ Long}
  \end{subfigure}
  \hspace{8pt}
  \begin{subfigure}{0.3\textwidth}
    \centering
    \setlength{\tabcolsep}{2pt}
    \begin{tabular}{@{}cc@{}}
      \includegraphics[width=0.48\linewidth]{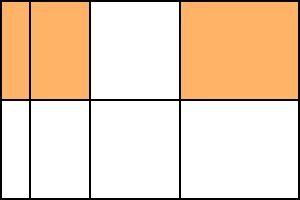} &
      \includegraphics[width=0.48\linewidth]{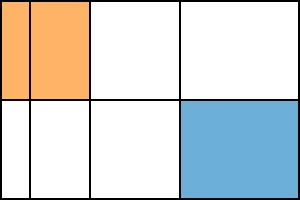} \\
    \end{tabular}
    \caption{Prehold vs.\ Preshift}
  \end{subfigure}

\caption{The three event types illustrated: hold vs.\ shift (given that person 1, orange, had been speaking prior to the window shown, whether they keep the floor or yield it), short vs.\ long (utterance duration), and prehold vs.\ preshift (activity preceding a hold/shift). Example configurations; other valid ones exist. }
  \label{fig:events}
\end{figure*}


\begin{table*}[h]
\centering
\small
\begin{tabular}{lcccc}
\toprule
\textbf{Model} & \textbf{Loss} & \textbf{H/S BA (\%)} & \textbf{Sh/L BA (\%)} & \textbf{Ph*/Ps BA (\%)} \\
\midrule
M0: Baseline                          & 2.166 & 76.41 & 84.11 & 63.90 \\
M1: Concatenation                     & 2.122 & 77.87 & 84.67 & 63.30 \\
M2: Cross-attention                   & 2.119 & 77.27 & 84.80 & 64.30 \\
M3: Cross-attention + $\Delta_3$      & 2.107 & 77.83 & 84.81 & 64.51 \\
M4: Cross-attention + $\Delta_{10}$   & 2.130 & 77.76 & 84.20 & 65.38 \\
M5: Cross-attention + $\Delta_3$ + Early Gates & \textbf{2.100} & 78.31 & \textbf{85.26} & 65.14 \\
M6: Cross-attention + $\Delta_3$ + Late Gates & 2.106 & \textbf{78.81} & 84.67 & \textbf{65.51} \\
\bottomrule
\end{tabular}
\smallskip
\begin{minipage}{\linewidth}
\small\textit{Note.} To assess statistical significance, we apply paired bootstrapping (1000 samples of 694 test conversations with replacement): \textbf{(M0 vs M5) 95\% CI}:  Loss: [-0.0691, -0.0616], H/S: [1.22, 2.60], S/L [0.59, 1.74], Ph/Ps [0.83, 2.03], i.e., all differences are significant. The asterisk refers to the subsample of Ph described in Section 4.3. 
\end{minipage}

\caption{Balanced accuracy (BA) and VAP loss for all models on the full test set. Bold indicates best performance per metric.}
\label{tab:models}
\end{table*}

\subsection{Multimodal Extensions}

We extend the baseline through three sequential design decisions: how to fuse the two modalities (M1, M2), how to add motion through delta features (M3, M4), and where to apply trainable gates (M5, M6). Each decision is a choice between two variants; we keep the lower-loss variant and build the next decision on top of it. Table~\ref{tab:models} shows the results of the configurations. 

\paragraph{Concatenation/Cross-attention} In M1 (concatenation), the visual feature vector (dimension 182) is concatenated with the audio representation (dimension 256) per speaker, then projected back to 256 dimensions via a linear layer:
\[
\tilde{x} = W_p [x ; v] + b_p, \quad W_p \in \mathbb{R}^{256 \times 438}.
\]
This output is then passed as an input to the audio-only architecture.  

In M2 (cross-attention), visual features are first linearly projected to dim 256, processed by self-attention, and then fused with audio via a two-step cross-attention. First, visual information is conditioned on audio:
\[
\tilde{X}_v = \text{CrossAttn}(X_v, X_a),
\]
then audio is updated using the audio-conditioned visual information:
\[
\tilde{X}_a = \text{CrossAttn}(X_a, \tilde{X}_v).
\]
The intuition is that audio-conditioned visual features provide a more informative key/value source for refining the audio representation.

\paragraph{Delta3/Delta10} Since the visual features encode static position rather than motion, we augment them with delta features:
\[
\Delta(\text{feat}_i)_j = (\text{feat}_i)_j - \frac{1}{N}\sum_{k=j-N}^{j-1} \text{feat}_k,
\]
capturing deviation from a short-term mean. We test $N=3$ (M3) and $N=10$ (M4). Adding delta features doubles the vector to dimension 364, which we map back to 256. 

\begin{figure}[h!]
  \centering
  \includegraphics[width=\columnwidth]{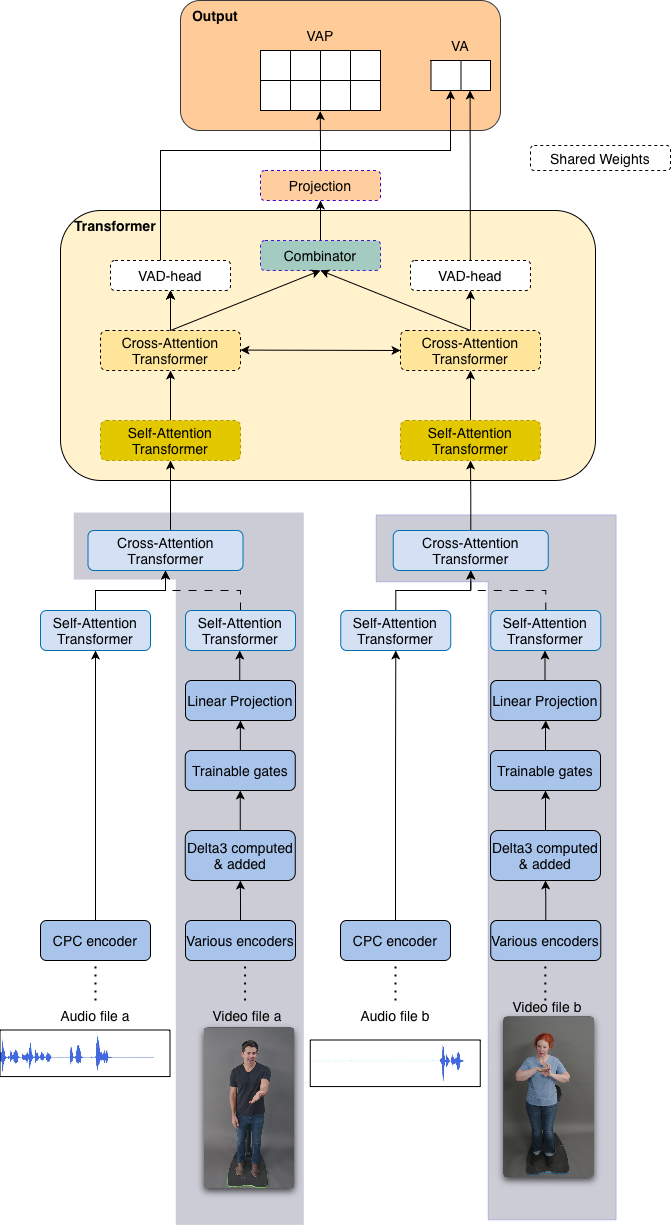}
    \caption{M5 architecture. Shaded blocks show components added to the M0 audio-only backbone, where two CPC-encoded streams pass per-channel self-attention, cross-speaker attention, and a combinator to predict VAP and VA. M5 adds visual features, augmented with deltas ($\Delta_3$), passed through early gates and self-attention, then cross-attended to audio before the main transformer.}
  \label{fig:M0-M5}
\end{figure}

\paragraph{Early/Late gates} A gate vector $g = \sigma(W_g x + b_g)$, with weights initialised to zero ($g = 0.5 ~ \forall x$ initially), scales features element-wise: $\bar{x} = g \odot x$. We try placing gates before the visual self-attention (early gates, M5) or before cross-attention (late gates, M6). The M5 model is shown in Figure \ref{fig:M0-M5}.

\subsection{Evaluation Metrics}

Beyond VAP loss, we evaluate on three interpretable turn-taking tasks \cite{ekstedt2022}. Examples of the events are shown in Figure \ref{fig:events}.

\textbf{Hold/Shift (H/S)}: accuracy at predicting whether a mutual silence ($\geq$100~ms, with 1~s pre/post single-speaker constraint) is followed by the current speaker continuing (Hold) or the other speaker taking over (Shift). This is perhaps the most central task for conversational systems to know whether the user is yielding the turn or not. 

\textbf{Short/Long (Sh/L)}: accuracy at predicting from the first 200~ms of a new speaker onset whether the utterance will be short ($<$1.5~s) or long ($\geq$1.5~s). For conversational systems, this can be useful to predict whether the user has just started a backchannel, in which case the system might continue, or is attempting to initiate a longer turn. 

\textbf{Shift Prediction (Ph/Ps)}: accuracy at predicting an upcoming turn shift while the current speaker is still active, measured as pre-shift (Ps) versus pre-hold (Ph). Due to heavy class imbalance, pre-hold events are subsampled to match pre-shift counts; we report the average over $N=10$
subsampling seeds. 


In addition to subsampling pre-holds, all tasks are evaluated using balanced accuracy (BA) to account for smaller class imbalances. 


Training used the Adam optimizer on an NVIDIA RTX 3090, the same learning rate as the baseline VAP, and early stopping with a patience of 5 epochs.

\section{Results}

\subsection{Model Comparison}

Table~\ref{tab:models} summarises results for the proposed architectures on the full test set, all using all features.

While the differences are not very big, every multimodal model reduces VAP loss compared to the baseline (M0: 2.166). M5 achieves the lowest loss (2.100) and is selected as the best architecture.

Cross-attention (M2) outperforms concatenation (M1) on loss and most metrics, though the difference is modest. Adding $\Delta_3$ features (M3) reduces loss further, indicating that motion information complements positional features. Gates (M5/M6) give further gains, as adaptive weighting helps the model suppress uninformative dimensions.

\subsection{Feature Group Ablation}

Using M5 architecture, we train separate models for each feature subset (Table~\ref{tab:features}). All Features achieve the best results overall, while FAU achieve the best result for the single feature groups. 

\begin{table}[ht]
\centering
\small
\begin{tabular}{lcccc}
\toprule
\textbf{Features} & \textbf{Loss} & \textbf{H/S} & \textbf{Sh/L} & \textbf{Ph*/Ps} \\
\midrule
Baseline    & 2.166 & 76.41 & 84.11 & 63.90 \\
All         & 2.100 & 78.31 & 85.26 & 65.14 \\
\hdashline
Body        & 2.143 & 75.51 & 83.97 & 63.64 \\
Gaze        & 2.150 & 76.18 & 84.03 & 62.49 \\
\textbf{FAU} & \textbf{2.115} & \textbf{77.52} & \textbf{84.89} & \textbf{63.95} \\
\bottomrule
\end{tabular}
\small\textit{Note.} The asterisk refers to the subsample of Ph. 
\caption{Feature group ablation using M5 architecture. Loss in absolute numbers balanced accuracy in \%.}
\label{tab:features}
\end{table}

In Table \ref{tab:CI's}, the differences between feature groups are assessed more closely. For FAU vs Body, three out of the four confidence intervals exclude zero; the only non-significant difference is for Ph/Ps. All four confidence intervals for FAU vs Gaze are significant. Both Body+FAU and Gaze+FAU outperform FAU alone on some metrics: Body+FAU has three CIs excluding zero, Gaze+FAU has two. This points to Body and Gaze providing complementary information when combined with FAU, even if their independent contribution is limited. It appears that body is particularly informative for H/S and Ph/Ps, and gaze for H/S. 


\begin{table}[ht]
\centering
\small 
\setlength{\tabcolsep}{4pt}
\resizebox{\columnwidth}{!}{%
\begin{tabular}{llcc}
\toprule
\textbf{Contrast} & \textbf{Metric} & \textbf{Mean $\Delta$} & \textbf{95\% CI} \\
\midrule
\multirow{4}{*}{FAU vs body}
& Loss  & $-0.028$ & $\textbf{[-0.032,\,-0.025]}$ \\
& H/S   & 2.01   & $\textbf{[1.27,\,2.66]}$ \\
& Sh/L  & 0.92   & $\textbf{[0.34,\,1.46]}$ \\
& Ph/Ps & 0.55   & $[-0.18,\,1.23]$ \\
\midrule
\multirow{4}{*}{FAU vs gaze}
& Loss  & $-0.035$ & $\textbf{[-0.038,\,-0.032]}$ \\
& H/S   & 1.34   & $\textbf{[0.65,\,2.05]}$ \\
& Sh/L  & 0.85   & $\textbf{[0.31,\,1.33]}$ \\
& Ph/Ps & 1.56   & $\textbf{[0.94,\,2.24]}$ \\
\midrule
\multirow{4}{*}{Body+FAU vs FAU}
& Loss  & $-0.009$ & $\textbf{[-0.010,\,-0.007]}$ \\
& H/S   & 0.74   & $\textbf{[0.22,\,1.29]}$ \\
& Sh/L  & 0.17   & $[-0.23,\,0.57]$ \\
& Ph/Ps & 1.00   & $\textbf{[0.43,\,1.56]}$ \\
\midrule
\multirow{4}{*}{Gaze+FAU vs FAU}
& Loss  & $-0.004$ & $\textbf{[-0.006,\,-0.001]}$ \\
& H/S   & 0.63   & $\textbf{[0.04,\,1.23]}$ \\
& Sh/L  & 0.18   & $[-0.28,\,0.63]$ \\
& Ph/Ps & $-0.38$  & $[-0.97,\,0.51]$ \\
\bottomrule
\end{tabular}%
}
\caption{Paired bootstrap differences for feature-complementarity contrasts. Balanced accuracy differences in percentage points while loss differences are absolute. For balanced accuracy, positive values indicate better performance by the first model in the contrast. For loss, negative values indicate lower loss for the first model. Confidence intervals that exclude zero are highlighted in bold. }
\label{tab:CI's}
\end{table}

\subsubsection{Individual FAU Importance}

We focus the finer-grained analysis on FAU because the feature-group comparison identified them as the strongest visual group overall (Tables~\ref{tab:features}-\ref{tab:CI's}). To estimate the contribution of individual action units, we zero out each FAU (and its delta) at inference and measure the resulting loss increase (Table~\ref{tab:fau}).

\begin{table}[ht]
\centering
\setlength{\tabcolsep}{4pt}
\resizebox{\columnwidth}{!}{%
\begin{tabular}{lcc}
\toprule
\textbf{Action Unit} & \textbf{Loss} & \textbf{Relative increase} \\
\midrule
None (full model) & 2.115 & -- \\
AU26 (JawDrop)    & 2.130 & $+0.71\%$ \\
AU4 (BrowLowerer) & 2.129 & $+0.66\%$ \\
AU43 (EyesClosed) & 2.126 & $+0.52\%$ \\
AU12 (LipCornerPull) & 2.124 & $+0.43\%$ \\
AU10 (UpperLipRaiser) & 2.122 & $+0.33\%$ \\
\midrule
\multicolumn{3}{l}{\textit{(remaining AUs: $\leq 0.14\%$ increase)}} \\
\bottomrule
\end{tabular}%
}
\caption{Loss increase when zeroing out single FAU. Top-5 most impactful shown.}
\label{tab:fau}
\end{table}

\begin{table*}[ht]
\centering
\resizebox{\textwidth}{!}{%
\begin{tabular}{lcccc}
\toprule
& \multicolumn{2}{c}{\textbf{P1}} & \multicolumn{2}{c}{\textbf{P2}} \\
\cmidrule(lr){2-3} \cmidrule(lr){4-5}
\textbf{AU} & Shift & Hold & Shift & Hold \\
\midrule
AU26 JawDrop & \textbf{0.058 [0.054, 0.062]} & \textbf{0.066 [0.064, 0.067]} & \textbf{0.048 [0.045, 0.051]} & \textbf{0.036 [0.035, 0.038]} \\
AU4 BrowLowerer & 0.367 [0.353, 0.380] & 0.359 [0.352, 0.366] & \textbf{0.363 [0.349, 0.377]} & \textbf{0.416 [0.409, 0.422]} \\
AU43 EyesClosed & \textbf{0.349 [0.332, 0.367]} & \textbf{0.394 [0.385, 0.403]} & 0.317 [0.301, 0.334] & 0.332 [0.324, 0.340] \\
AU12 LipCornerPull & \textbf{0.525 [0.507, 0.543]} & \textbf{0.390 [0.383, 0.397]} & \textbf{0.511 [0.494, 0.529]} & \textbf{0.455 [0.448, 0.463]} \\
AU10 UpperLipRaiser & 0.352 [0.338, 0.365] & 0.350 [0.344, 0.356] & \textbf{0.363 [0.350, 0.376]} & \textbf{0.275 [0.270, 0.281]} \\
\bottomrule
\end{tabular}%
}
\caption{Action unit averages, with confidence intervals, for the last frame before each event, for the active participant 1 (P1) and the passive participant 2 (P2) under Shift and Hold conditions.}
\label{tab:au_values}
\end{table*}

JawDrop (AU26), BrowLowerer (AU4), and EyesClosed (AU43) are the most important individual units. These results are intuitively plausible: jaw movement is directly tied to speaking, and eye closure relates to timing and emphasis. However, we acknowledge that the FAU are likely highly correlated, and that this approach only shows influence when one is zeroed at a time. 

Table \ref{tab:au_values} shows the value for each of the most influential FAU for the frame right before a hold/shift event, together with its $95\%$ confidence interval, where P1 is the active speaker. It indicates that P1 is more likely to close their eyes (AU43) and have their mouth open (AU26) when holding the turn, rather than yielding it. Also, P1 is more likely to smile (AU12) when yielding the turn. P2 is more likely to open their mouth (AU10, AU26) and smile (AU12) when they are about to take the turn, while they tend to lower their eyebrows (AU4) when the other participant holds the turn.

\begin{table}[htbp]
\centering
\small
\begin{tabular}{lccc}
\toprule
Domain & Shift/min & Hold/min & Shift/Hold \\
\midrule
Improvised & 1.10 & 4.68 & 0.24 \\
Naturalistic & 1.41 & 6.56 & 0.21 \\
\bottomrule
\end{tabular}
\caption{Shift and hold rates by domain per minute, and Shift/Hold ratio. }
\label{tab:domain_stats}
\end{table}

\begin{table*}[h]
\centering
\small
\setlength{\tabcolsep}{2.5pt}
\begin{tabular*}{\textwidth}{@{\extracolsep{\fill}}llcccccccccc}
\toprule
& & 
& \multicolumn{3}{c}{Hold/Shift}
& \multicolumn{3}{c}{Short/Long}
& \multicolumn{3}{c}{Pre-Hold/Pre-Shift} \\
\cmidrule(lr){4-6}\cmidrule(lr){7-9}\cmidrule(lr){10-12}
Train & Test
& Loss
& BA & H & S
& BA & Sh & L
& BA & Ph* & Ps \\
\midrule
Imp & Imp & 2.235 & 79.41 & \textbf{87.20} & \textbf{71.63} & 84.45 & \textbf{73.52} & \textbf{95.39} & 66.46 & \textbf{94.25} & \textbf{38.66} \\
Imp & Nat & 2.044 & 77.75 & \textbf{84.05} & \textbf{71.45} & 80.61 & \textbf{64.66} & \textbf{96.55} & 66.80 & \textbf{93.16} & \textbf{40.44} \\
Nat & Imp & 2.304 & 77.80 & \textbf{89.37} & \textbf{66.23} & 84.50 & \textbf{77.42} & \textbf{91.58} & 64.63 & \textbf{95.19} & \textbf{34.07} \\
Nat & Nat & 2.009 & 77.04 & \textbf{88.42} & \textbf{65.65} & 82.15 & \textbf{70.53} & \textbf{93.77} & 64.98 & \textbf{96.09} & \textbf{33.87} \\
\bottomrule
\end{tabular*}
\small\textit{Note.} The asterisk refers to the subsample of Ph described in Section 4.3. 
\caption{Cross-testing results between improvised and naturalistic training and test sets. BA denotes balanced accuracy.}
\label{tab:imp-nat-hs}
\end{table*}

\subsection{Improvised vs. Naturalistic}

The M5 model (all features) achieves a lower, i.e. better, VAP loss on naturalistic conversations than on improvised ones ($1.968$ vs.\ $2.203$). The pattern reverses for the event-based metrics: improvised conversations yield higher balanced accuracy on hold/shift ($78.86\%$ vs.\ $77.89\%$) and short/long ($85.71\%$ vs.\ $83.90\%$), while the pre-hold/pre-shift difference is small. We assess significance using stratified bootstrapping, resampling improvised and naturalistic subsets separately and computing the improvised$-$naturalistic difference. Only the loss difference is significant ($\Delta = 0.235$, CI $[0.148, 0.324]$), confirming that the model predicts naturalistic dynamics more accurately overall. The balanced-accuracy differences --- H/S ($\Delta = 0.88$, CI $[-1.08, 2.75]$), S/L ($\Delta = 1.73$, CI $[-0.42, 3.96]$), and Ph/Ps ($\Delta = 0.14$, CI $[-2.19, 2.55]$) --- all have CIs spanning zero, so the trend toward higher event accuracies on improvised data is not significant. One explanation is that improvised conversations have more exaggerated turn-taking cues, making discrete events easier to classify, whereas naturalistic conversations have smoother dynamics that the model captures more accurately at the level of overall voice-activity distribution. This remains an open question for future work. 

Table \ref{tab:domain_stats} shows the shift and holds in the naturalistic and improvised test-sets. Naturalistic conversations show more shifts and holds per minute, indicating more frequent pauses, whereas speakers in improvised conversation produce longer interpausal units.

Table \ref{tab:imp-nat-hs} shows how models perform on partitions of the test set, when trained on naturalistic or improvised. Models trained on improvised data appear to perform better on shift, while the models trained on naturalistic appear to perform better on hold. This might imply that the improvised models lean towards predicting shift while the naturalistic models tend to predict hold, disregarding test-set. The same directional bias appears on the other tasks. On pre-hold/pre-shift, improvised-trained models favour pre-shift while naturalistic-trained models favour pre-hold, which follows directly from the shift/hold ratio (Table~\ref{tab:domain_stats}). On short/long, improvised-trained models are prone to predict long, while naturalistic-trained models favour short; this is consistent with the same account, although the underlying short/long base rates are not reported here. Both effects are smaller than for hold/shift. We present these as descriptive trends and do not draw significance-based conclusions from them. 

\section{Discussion}

Every multimodal variant lowers VAP loss relative to the audio-only baseline (Table~\ref{tab:models}), and for our best model the gains are statistically significant across loss and all three pre-defined tasks. This confirms that visual features add information that audio alone does not capture. 

Concatenation (M1) and cross-attention (M2) perform almost identically on loss and on the event-based metrics. We anticipated that cross-attention, with its explicit conditioning between modalities, would show a clearer advantage. One interpretation is that the audio transformer integrates the visual features regardless of how they enter the network; another is that at this data scale the fusion mechanism matters less than the presence of the visual signal itself. We cannot separate these explanations from the present experiments.


Among the feature groups, FAU are the most informative. Alone, they come close to the performance of the all features model and clearly outperform body and gaze (Table~\ref{tab:features}), a gap that is significant for loss and for the hold/shift and short/long tasks (Table~\ref{tab:CI's}). The same ranking has been reported for mediated and video-conferencing data \cite{onishi2023,oconnor2025}; our results show that it also holds for large-scale face-to-face conversation. Body and gaze add little in isolation, yet the full model still improves over any single group, so these features carry information that is complementary rather than redundant.


The size of the gains is itself informative. On clean, channel-separated audio the acoustic signal already carries most of what the task requires, so visual features are better understood as supplementing audio than replacing it. They help most where the acoustic evidence is ambiguous, which fits the finding that facial action units are the strongest visual contributor: jaw, mouth, eye, and brow movement are tied directly to speech onset and offset and to communicative intent. This also implies that the gains reported here may be conservative: visual access to the face and lips raises speech intelligibility in noise \cite{sumby1954}, and a turn-taking model evaluated under added noise loses much of its audio-only hold/shift accuracy while a multimodal model degrades far less \cite{oconnorrussell2025noise}. The recordings used are clean and the two speakers occupy separate channels, so the contribution of visual features may be larger under noisier, single-channel conditions typical of deployed dialogue systems.

\section{Conclusion}

Visual features from large-scale face-to-face conversations improve turn-taking prediction over the audio-only VAP. Our best model (M5: cross-attention with $\Delta_3$ features and early gates) shows statistically significant improvements across all three evaluation tasks. Among individual feature groups, FAU are the most informative, though combining all features performs best. Also, the results appear to vary if trained or tested on acted or non-acted conversations. These results support the use of visual information in spoken dialogue systems aiming for more natural turn-taking behaviour.

\section{Limitations}

Several limitations qualify these results. Gaze estimates are unreliable when the eyes are partly occluded, which likely caps the contribution of the gaze group. A single averaging window $N$ for the delta features is unlikely to suit feature types with different temporal dynamics. The multimodal model is also more costly to run than the audio-only baseline, a consideration for online use \cite{inoue2024realtime}. Promising directions include end-to-end video encoders that remove the need for hand-engineered feature extraction, and per-group tuning of the delta window $N$.

A final limitation concerns speaker and dyad identity. Turn-taking timing is highly dyad-specific: \citet{cavalcanti2025} find that idiosyncratic and ``dyadosyncratic'' factors dominate variation in transition timing. This affects our setup in two ways. Our splits are conversation-disjoint but not dyad-disjoint, so the model may exploit dyad-specific regularities seen in training; the reported scores should thus be read as an upper bound, pending a dyad-disjoint evaluation. Because the model is trained pooled across dyads with no speaker or dyad conditioning, its accuracies are averages over a heterogeneous population. Conditioning on, or adapting to, the speaker or dyad is a natural direction for future work.

\section{Acknowledgements}

This study was financed, in part, by the São Paulo Research Foundation (FAPESP), Brazil. Grants \#2024/06797-2, and \#22/05848-7. 

\bibliography{custom}

\end{document}